\documentclass[a4paper,11pt]{article}
\usepackage{pos}
\usepackage{subfig}

\title{Latest view of CTA 1 with VERITAS}
\author*[a]{Alisha Chromey}
\onbehalf{for the VERITAS Collaboration}

\affiliation[a]{Center for Astrophysics | Harvard \& Smithsonian, 60 Garden Street, Cambridge, MA 021138, USA}

\emailAdd{alisha.chromey@cfa.harvard.edu}

\abstract{CTA 1 is a shell-type supernova remnant (SNR) with a central pulsar wind nebula (PWN), visible at very-high-energy (VHE) from 50 GeV to 100 TeV from a moderately extended emission region. While general consensus concludes the VHE emission originates from relativistic leptons accelerated by the PWN and undergoing inverse Compton scattering, questions remain about electron escape and propagation, as well as the evolutionary stage of this particular PWN. CTA 1 is on the cusp of middle age ($\sim$13 kyr) and spatially resolvable at energies visible to imaging atmospheric Cherenkov telescopes (IACTs), such as the Very Energetic Radiation Imaging Telescope Array System (VERITAS) (PSF < 0.1 deg).  Therefore, this remnant is an excellent candidate to study lepton propagation and escape between different PWN evolutionary stages. Since the initial VERITAS publication on CTA 1 in 2013, VERITAS has performed new observations, adding to a total exposure of about 120 hours. We have analyzed the entire VERITAS CTA 1 dataset to date and report results.}

\ConferenceLogo{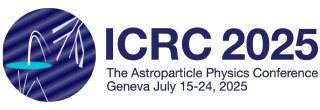}

\FullConference{39th International Cosmic Ray Conference (ICRC2025)\\
 15–24 July 2025\\
Geneva, Switzerland\\}

\begin{document}
\maketitle

\section{Introduction}
The post-core-collapse relics of massive stars are candidates for astrophysical accelerators of galactic-sourced cosmic rays, and pulsar wind nebulae (PWNe) are known sources of accelerated electrons and positrons \cite{pulsarpaper}. However, many details of cosmic ray transport remain unknown; the energies of the accelerated particles and transport mechanisms are greatly dependent on the evolving PWN and interactions with surroundings \cite{pwnlit}. Spectral and morphological studies on several PWN, across a diverse sample of evolutionary states and environments, is the path forward to elucidate details of cosmic ray production. Indirect measurements of leptonic interactions within environments of astrophysical accelerators are possible by observing neutral particle products, such as $\gamma$-rays produced at very-high-energies (VHE).

CTA 1 is a composite shell-type supernova remnant (SNR) and pulsar wind nebula, located at a distance of $1.09\pm0.2$ kpc \cite{doetal}. The incomplete shell is visible in radio, whereas a pulsar coincident and associated with the SNR, RX J0007.0+7303, is detected only in X-rays and $\gamma$-rays. The age of the pulsar is $\sim$13 kyr \cite{caraveo}\cite{lin}, consistent with estimated age of the SNR \cite{slane}, placing it either as a late stage free-expansion or early stage reverberation. This calculated age, combined with CTA 1's relatively high galactic latitude ($+10.4^{\circ}$) and spatially resolvable features across multiple wavebands make it a good candidate to study leptonic transport and PWN evolution and interaction with the SNR shell.

Multiple instruments have detected extended $\gamma$-ray emission associated with CTA 1. In 2013, the VERITAS collaboration published spectral and morphological results on about 41 hours of observations, detecting TeV $\gamma$-ray emission at 6.5$\sigma$ post-trials with an extended morphology modeled by an elliptical 2D Gaussian, 0.30$^{\circ}$ $\pm$ 0.03$^{\circ}$(0.24$^{\circ}$ $\pm$ 0.03$^{\circ}$) semi-major(minor) axes. The spectrum, normalized at 3 TeV, is best characterized by a power-law with differential spectral index $\Gamma=2.2\pm0.2_{stat}\pm0.3_{sys}$ and $N_{0}=(9.1\pm1.3_{stat}\pm1.7_{sys}) \times 10^{-14}cm^{-2}s^{-1}TeV^{-1}$ \cite{veritascta}. In 2024 \textit{Fermi}-LAT published results from about 15 years of data, claiming possible detection of extended $\gamma$-ray emission associated with the CTA 1 PWN. The analysis was performed from 50 GeV to 1 TeV in order to remove the effect of the bright $\gamma$-ray pulsar \cite{fermicta}. In its first catalog LHAASO reported extended TeV emission associated with CTA 1 \cite{lhaasocat}. In 2024 LHAASO published a dedicated morphological and spectral study on the CTA 1 region, reporting 21$\sigma$ at 8-100 TeV and 17$\sigma$ >100 TeV and extensions 0.23$^{\circ}$ $\pm$ -0.03$^{\circ}$ and 0.17$^{\circ}$ $\pm$ 0.03$^{\circ}$, respectively \cite{lhaasocta}. All three studies claim that the VHE emission originates from relativistic electrons accelerated in the PWN.

\section{VERITAS}

VERITAS is an array of four 12-meter imaging atmospheric Cherenkov telescopes located at the Fred Lawrence Whipple Observatory (FLWO) in southern Arizona (31 40N, 110 57W,  1.3km a.s.l.). Each telescope dish consists of 345 facets and operates with pointing accuracy < 50 arcsecs. The array is sensitive from 100 GeV to 30 TeV, with an energy resolution between 15-25\%. The FOV is 3.5 deg and the 68\% containment is <0.1 deg at 1 TeV. The array can detect 1\% Crab VHE flux in $\sim$25 hours at high elevation angles. For full details of VERITAS and its performance see \cite{nahee} and \cite{throughput}.

\section{Observations}
VERITAS observed CTA 1 from September 2010 to January 2011 and then from September to December 2011. The 2013 publication is derived from this data set. About 38 hours of observations from the 2013 dataset are included in the analysis presented in this proceeding.

In order to achieve a deeper exposure, further observations of the CTA 1 region were taken from September 2023 to January 2024 and then September to December 2024. All VERITAS data were taken in wobble mode, pointing 0.7$^{\circ}$ offset from the pulsar position. The scientific motivations for a larger dataset are an energy dependent morphology study and a spectral analysis improved over the 2013 studies with greater statistics at the lowest and highest energy bins.

\section{Analysis and Discussion}
A series of quality selection cuts are performed on the data. Only four-telescope data are included. After correcting for deadtime the entire dataset, collected before 2013 and afterwards from 2023 to 2024, adds up to an exposure of 118.6 hours. Analysis is performed using an internal analysis software package, VEGAS \cite{vegas}. The reflected region method of estimating background counts in off-source regions is used to produce an energy spectrum. The ring background method of estimating background counts is used to produce a 2D significance map of the source region \cite{berge}. In order to reduce the cosmic-ray background a series of gamma-hadron separation cuts at moderate levels are applied during data reduction steps. This analysis adds the Image Template Method (ITM), using template predictions and maximizing the likelihood during event reconstruction, which improves angular and energy reconstruction \cite{itm}. The on-source counts are derived from a circular region r=0.25$^{\circ}$ and calculation of background is excluded from a region r=0.75$^{\circ}$, both centered around the position of the previously reported VERITAS centroid. To crosscheck results, a secondary analysis was performed on the same dataset with a separate analysis package, EventDisplay \cite{ed}.

\begin{figure}
    \centering
    \includegraphics[width=.45\linewidth]{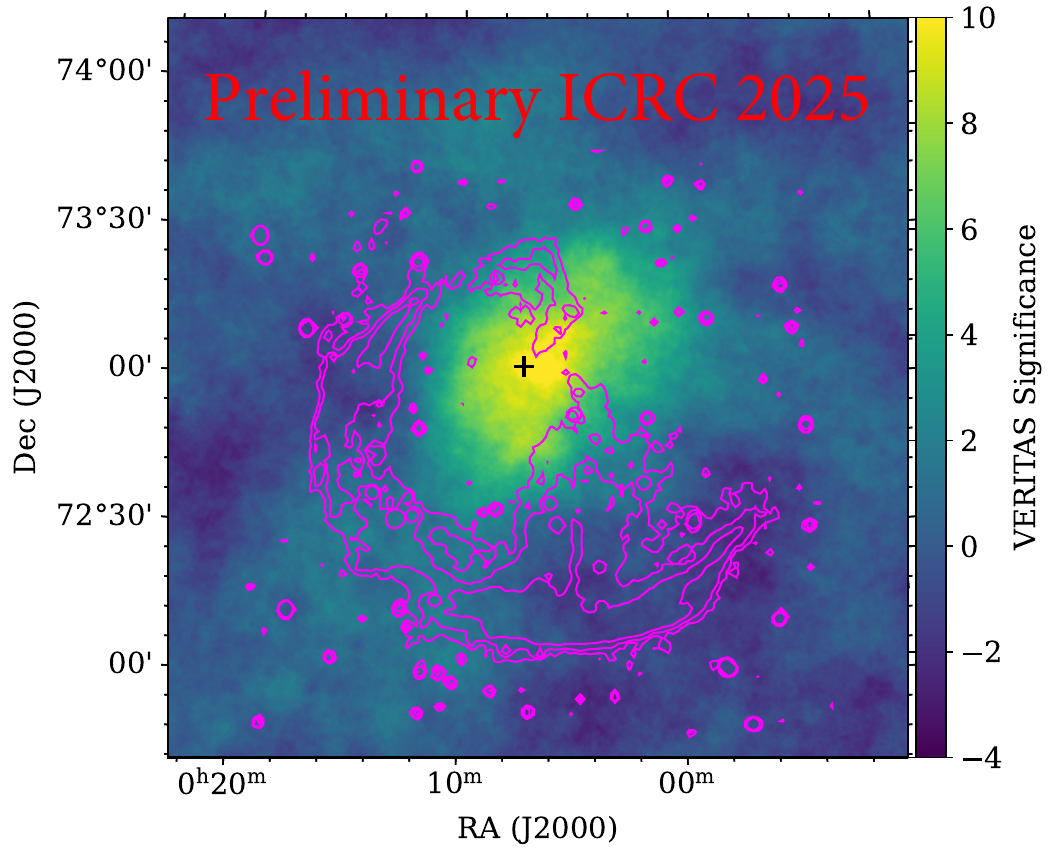}
    \includegraphics[width=.45\linewidth]{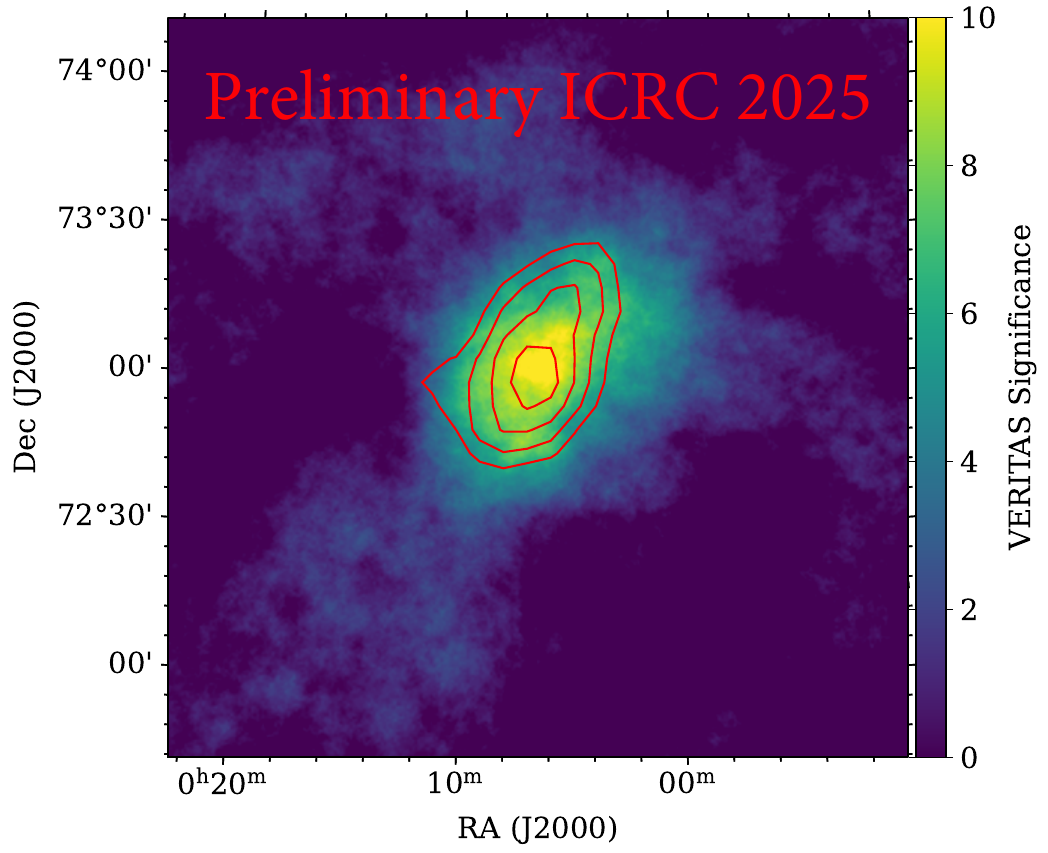}
    \caption{
    \small
    (Left) The VERITAS 2D significance map with 1420 MHz contours overlay \cite{radiocontours} and the pulsar position denoted by a black cross. (Right) The same VERITAS 2D significance map with the lower limit of the color scale set at zero and 2013 VERITAS significance contours (3, 4, 5, and 6$\sigma$) overlay.
    }
    \label{fig:sig_maps}
\end{figure}

Analysis at the 2013 VERITAS source, VER J0006+729, produces excess counts with significance 9.1$\sigma$. Figure 1 presents VERITAS significance maps of the CTA 1 region. The $\gamma$-ray emission is clearly extended, overlapping the pulsar position, and consistent with the previous VERITAS published source extension. The VHE photon spectrum is well fit with a power-law at normalized energy 3 TeV. The low energy threshold is $\sim$500 GeV and the integrated flux above 1 TeV is $(6.88\pm0.92_{stat}) \times 10^{13} cm^{-2}s^{-1}$.

Events are separated above and below 3.0 TeV and separately analyzed with the ring background method. The locations of highest significance from the analyses and the position of the pulsar associated with CTA1 are listed together in Table 1.


\begin{table} \centering
\caption{The position of pulsar RX J0007.0+7303 listed with the max significance positions of $\gamma$-ray events above and below 3 TeV. The systematic uncertainty of the position due to the VERITAS array pointing error is $\sim$50 arc-seconds.}
\label{data}
\begin{tabular}{l|cc} \hline
\textbf{} & \textbf{RA deg} (J2000) & \textbf{Dec deg} (J2000) \\ \hline
Pulsar & 1.78 & 73.05 \\
E > 3 TeV & 1.74 & 73.05 \\
E < 3 TeV & 1.53 & 73.12 \\ \hline
\end{tabular}
\end{table}

\section{Summary and Conclusion}

The VERITAS collaboration carried out a deeper exposure on the CTA 1 region, more than doubling the dataset employed in the VERITAS 2013 discovery publication. A detailed analysis for spectral and energy-dependent morphological studies is underway, with preliminary results presented in this proceeding. A broadband VHE spectrum, with PWN models, and additional morphological studies will be presented in an upcoming published paper.

\section{Acknowledgments}

This research is supported by grants from the U.S. Department of Energy Office of Science, the U.S. National Science Foundation and the Smithsonian Institution, by NSERC in Canada, and by the Helmholtz Association in Germany. This research used resources provided by the Open Science Grid, which is supported by the National Science Foundation and the U.S. Department of Energy's Office of Science, and resources of the National Energy Research Scientific Computing Center (NERSC), a U.S. Department of Energy Office of Science User Facility operated under Contract No. DE-AC02-05CH11231. We acknowledge the excellent work of the technical support staff at the Fred Lawrence Whipple Observatory and at the collaborating institutions in the construction and operation of the instrument.

\section*{Full Author List: VERITAS Collaboration}

\scriptsize
\noindent
A.~Archer$^{1}$,
P.~Bangale$^{2}$,
J.~T.~Bartkoske$^{3}$,
W.~Benbow$^{4}$,
Y.~Chen$^{5}$,
J.~L.~Christiansen$^{6}$,
A.~J.~Chromey$^{4}$,
A.~Duerr$^{3}$,
M.~Errando$^{7}$,
M.~Escobar~Godoy$^{8}$,
J.~Escudero Pedrosa$^{4}$,
Q.~Feng$^{3}$,
S.~Filbert$^{3}$,
L.~Fortson$^{9}$,
A.~Furniss$^{8}$,
W.~Hanlon$^{4}$,
O.~Hervet$^{8}$,
C.~E.~Hinrichs$^{4,10}$,
J.~Holder$^{11}$,
T.~B.~Humensky$^{12,13}$,
M.~Iskakova$^{7}$,
W.~Jin$^{5}$,
M.~N.~Johnson$^{8}$,
E.~Joshi$^{14}$,
M.~Kertzman$^{1}$,
M.~Kherlakian$^{15}$,
D.~Kieda$^{3}$,
T.~K.~Kleiner$^{14}$,
N.~Korzoun$^{11}$,
S.~Kumar$^{12}$,
M.~J.~Lang$^{16}$,
M.~Lundy$^{17}$,
G.~Maier$^{14}$,
C.~E~McGrath$^{18}$,
P.~Moriarty$^{16}$,
R.~Mukherjee$^{19}$,
W.~Ning$^{5}$,
R.~A.~Ong$^{5}$,
A.~Pandey$^{3}$,
M.~Pohl$^{20,14}$,
E.~Pueschel$^{15}$,
J.~Quinn$^{18}$,
P.~L.~Rabinowitz$^{7}$,
K.~Ragan$^{17}$,
P.~T.~Reynolds$^{21}$,
D.~Ribeiro$^{9}$,
E.~Roache$^{4}$,
I.~Sadeh$^{14}$,
L.~Saha$^{4}$,
H.~Salzmann$^{8}$,
M.~Santander$^{22}$,
G.~H.~Sembroski$^{23}$,
B.~Shen$^{12}$,
M.~Splettstoesser$^{8}$,
A.~K.~Talluri$^{9}$,
S.~Tandon$^{19}$,
J.~V.~Tucci$^{24}$,
J.~Valverde$^{25,13}$,
V.~V.~Vassiliev$^{5}$,
D.~A.~Williams$^{8}$,
S.~L.~Wong$^{17}$,
and
T.~Yoshikoshi$^{26}$\\
\\
\noindent
$^{1}${Department of Physics and Astronomy, DePauw University, Greencastle, IN 46135-0037, USA}

\noindent
$^{2}${Department of Physics, Temple University, Philadelphia, PA 19122, USA}

\noindent
$^{3}${Department of Physics and Astronomy, University of Utah, Salt Lake City, UT 84112, USA}

\noindent
$^{4}${Center for Astrophysics $|$ Harvard \& Smithsonian, Cambridge, MA 02138, USA}

\noindent
$^{5}${Department of Physics and Astronomy, University of California, Los Angeles, CA 90095, USA}

\noindent
$^{6}${Physics Department, California Polytechnic State University, San Luis Obispo, CA 94307, USA}

\noindent
$^{7}${Department of Physics, Washington University, St. Louis, MO 63130, USA}

\noindent
$^{8}${Santa Cruz Institute for Particle Physics and Department of Physics, University of California, Santa Cruz, CA 95064, USA}

\noindent
$^{9}${School of Physics and Astronomy, University of Minnesota, Minneapolis, MN 55455, USA}

\noindent
$^{10}${Department of Physics and Astronomy, Dartmouth College, 6127 Wilder Laboratory, Hanover, NH 03755 USA}

\noindent
$^{11}${Department of Physics and Astronomy and the Bartol Research Institute, University of Delaware, Newark, DE 19716, USA}

\noindent
$^{12}${Department of Physics, University of Maryland, College Park, MD, USA }

\noindent
$^{13}${NASA GSFC, Greenbelt, MD 20771, USA}

\noindent
$^{14}${DESY, Platanenallee 6, 15738 Zeuthen, Germany}

\noindent
$^{15}${Fakult\"at f\"ur Physik \& Astronomie, Ruhr-Universit\"at Bochum, D-44780 Bochum, Germany}

\noindent
$^{16}${School of Natural Sciences, University of Galway, University Road, Galway, H91 TK33, Ireland}

\noindent
$^{17}${Physics Department, McGill University, Montreal, QC H3A 2T8, Canada}

\noindent
$^{18}${School of Physics, University College Dublin, Belfield, Dublin 4, Ireland}

\noindent
$^{19}${Department of Physics and Astronomy, Barnard College, Columbia University, NY 10027, USA}

\noindent
$^{20}${Institute of Physics and Astronomy, University of Potsdam, 14476 Potsdam-Golm, Germany}

\noindent
$^{21}${Department of Physical Sciences, Munster Technological University, Bishopstown, Cork, T12 P928, Ireland}

\noindent
$^{22}${Department of Physics and Astronomy, University of Alabama, Tuscaloosa, AL 35487, USA}

\noindent
$^{23}${Department of Physics and Astronomy, Purdue University, West Lafayette, IN 47907, USA}

\noindent
$^{24}${Department of Physics, Indiana University Indianapolis, Indianapolis, Indiana 46202, USA}

\noindent
$^{25}${Department of Physics, University of Maryland, Baltimore County, Baltimore MD 21250, USA}

\noindent
$^{26}${Institute for Cosmic Ray Research, University of Tokyo, 5-1-5, Kashiwa-no-ha, Kashiwa, Chiba 277-8582, Japan}

\end{document}